\documentclass[11pt]{article}
\pdfoutput=1 
\usepackage{jheppub}
\usepackage{graphicx,verbatim}
\usepackage{psfrag, times}
\usepackage{upgreek}
\usepackage[T1]{fontenc}
\usepackage[vcentermath]{youngtab}

\def\cN{{\mathcal{N}}}

\def\bbR{{\mathbb{R}}}
\def\bC{{\mathbb{C}}}

\def\bz{{\bar{z}}}
\def\cE{{\mathcal{E}}}
\def\cH{{\mathcal{H}}}

\def\cO{{\mathcal{O}}}
\def\cG{{\mathcal{G}}}
\def\nn{\nonumber}
\def\cF{{\mathcal{F}}}
\def\cL{{\mathcal{L}}}
\def\be{\begin{equation}}
\def\ee{\end{equation}}

\def\bea{\begin{eqnarray}}
\def\eea{\end{eqnarray}}

\def\Del{\Delta}

\def\lam{\lambda}
\def\veps{\varepsilon}

\def\Fpm{ F_{\lam_+\lam_-}^{(\veps)}}

\def\cFpm{ {\mathcal{F}}_{\lam_+\lam_-}}
\def\psipm{\psi_{\lam_+\lam_-}^{(\veps)}}

\def\ru{{\rm{u}}}
\def\rv{{\rm{v}}}

\def\bu{{\bar{u}}}
\def\rK{{\rm K}}
\def\Jueps{J^{(\varepsilon)}_u}
\def\Jbueps{J^{(\varepsilon)}_\bu}
\def\Ju0{J^{(0)}_u}
\def\Jbu0{J^{(0)}_\bu}
\def\ba{{\bf{a}}}
\def\bb'{{\bf{b'}}}
\def\bb{{\bf{b}}}
\def\cI{{\mathcal{I}}}
\def\trK{\tilde{\rm{K}}}

\def\tveps{t^{(\varepsilon)}}
\def\hcI{\hat{\cI}}
\def\hJueps{\hat{J}^{(\varepsilon)}_u}
\def\hJbueps{\hat{J}^{(\varepsilon)}_\bu}
\def\hJu0{\hat{J}^{(0)}_u}
\def\hJbu0{\hat{J}^{(0)}_\bu}
\def\hJua{\hat{J}^{(\ba)}_u}
\def\hJbua{\hat{J}^{(\ba)}_\bu}
\def\hLup{\hat{L}_{u\bu}^{(+)}}
\def\hLum{\hat{L}_{u\bu}^{(-)}}
\def\eps{\epsilon}
\def\bI{{\bf{I}}}
\def\bJ{{\bf{J}}}

\begin{document}
\title{\Large \bf Quantum Integrable Systems from Conformal Blocks }
\author[]{Heng-Yu Chen${}^{1}$ and Joshua D. Qualls${}^{1}$}
\affiliation{$^1$Department of Physics and Center for Theoretical Sciences, \\
National Taiwan University, Taipei 10617, Taiwan}
\emailAdd{ heng.yu.chen@phys.ntu.edu.tw,  joshqualls@ntu.edu.tw  } 
\vspace{2cm}
\abstract
{
In this note, we extend the striking connections between quantum integrable systems and conformal blocks recently found in \cite{Isachenkov:2016gim} in several directions.
First, we explicitly demonstrate that the action of quartic conformal Casimir operator on general d-dimensional scalar conformal blocks,
can be expressed in terms of certain combinations of commuting integrals of motions of two particle hyperbolic ${\rm BC}_2$ Calogero-Sutherland system. 
The permutation and reflection properties of the underlying Dunkl operators play crucial roles in establishing such a connection. 
Next, we show that the scalar superconformal blocks in SCFTs with four and eight supercharges and suitable chirality constraints 
can also be identified with the eigenfunctions of the same Calogero-Sutherland system, this demonstrates the universality of such a connection.
Finally, we observe that the so-called ``seed'' conformal blocks for constructing four point functions for operators with arbitrary space-time spins
in four dimensional CFTs can also be linearly expanded in terms of Calogero-Sutherland eigenfunctions.
 }
\maketitle

\section{Introduction and Summary}
\paragraph{}
Recently, a striking new connection between conformal field theories and quantum integrable systems has been proposed in \cite{Isachenkov:2016gim}.
The authors mapped the action of the conformal quadratic Casimir on the conformal block for $d$-dimensional scalar primary fields, via a certain coordinate transformation, 
to the Hamiltonian of a variant of the hyperbolic Calogero-Sutherland system \cite{Calogero:1970nt, Sutherland}.
As a result, we can identify the scalar conformal blocks, 
which were previously known in an explicit form only in even dimensions in terms of hypergeometric functions \cite{Dolan:2003hv, Dolan:2011dv},
with the eigenfunctions of the corresponding quantum integrable system. 
These eigenfunctions have been studied and can be identified with the degenerate limit of so-called virtual Koornwinder polynomials \cite{Rains:2005};
this extends the story relating symmetric Jack polynomials with the exact eigenfunction of  trigonometric Calogero-Sutherland model (See e.g. \cite{Lapointe:1995ap} for a nice exposition).
These exact expression for the conformal blocks provide the important elements for constructing various correlation functions in arbitrary dimensions.
\paragraph{}
It is well-known that in addition to the Hamiltonian, a quantum integrable system possesses a tower of algebraically independent commuting integrals of motion. For quantum Calogero-Sutherland systems, these commuting objects can be most conveniently expressed as products of commuting Dunkl operators \cite{Dunkl:1989} which are typically defined on discrete lattice systems.
While the Hamiltonian/quadratic Casimir with appropriate boundary conditions is sufficient for finding the conformal blocks, in light of this new relation it is natural to ask if the extra commuting integrals of motion defining the quantum integrable system also have counterparts in conformal field theories.
The answer is affirmative, and as somewhat expected they are mapped to the combination of quadratic and other Casimir operators which can be constructed from both conformal and superconformal generators\footnote{The quantum integrable system emerging here is sufficiently simple that it only has two degrees of freedom; thus only two independent commuting integrals of motion can be constructed.}. 
To firmly establish the intuitive statement that commuting conformal Casimir operators correspond to commuting integrals of motion, we must identify the precise combination of conformal Casimir operators. As we will demonstrate in this note, however, in order to achieve this it is crucial to keep track of the various reflection and permutation operators within the Dunkl operators.
\paragraph{}
In this note we also extend the connections with quantum integrable systems to a special class of superconformal blocks in superconformal field theories with four and eight supercharges.
This follows from two simple observations. First, superconformal blocks involving chiral primary scalars can be linearly expanded in terms of conformal blocks for primary scalars \cite{Poland:2010wg, N=1 Blocks}, moreover these conformal blocks themselves involve various recurrence relations \cite{Dolan:2003hv, Dolan:2011dv}. 
Together, these facts lead to multiplicative relations between conformal and superconformal blocks observed also earlier in \cite{Bobev:2015jxa}, 
which can now be regarded as part of the coordinate transformation leading to the quantum integrable systems. 
We will show explicitly that the same variant of hyperbolic Calogero-Sutherland system can also be connected to these superconformal blocks,
with appropriate shifts in the coefficients which are consistent with the multiplicative relations. 
\paragraph{}
We also discuss how this connection between conformal field theory and quantum integrable systems can be extended to correlation function having external fields with nonzero space-time spins\footnote{For a nice recent work on the action of quadratic Casimir on conformal block for operators with spins, see \cite{Echeverri:2016dun}. }. In this note, we consider how seed blocks in four spacetime dimensions translate into Hilbert space ``seeds''. 
These Hilbert space seeds are expressed in terms of Calogero-Sutherland eigenfunctions through a certain diagonalization procedure. 
Combining this observation with the analysis for superconformal blocks, scalar conformal blocks or equivalently the eigenfunctions of hyperbolic Calogero-Sutherthand Hamiltonian appear to be the most natural basis for expressing various other correlation functions. 
We speculate how similar relation holds in other space-time dimensions.
\paragraph{}
Let us begin by briefly reviewing the main results in \cite{Isachenkov:2016gim} and specifying our notations.
For a non-SUSY CFT in $d$-dimensions, the four point function for primary scalar operators $\phi_i(x)$ with scale dimensions $\Delta_i$ takes the form \cite{Dolan:2011dv}:
\be\label{4pt-function-Para}
\langle \phi_1(x_1)\phi_2(x_2)\phi_3(x_3)\phi_4(x_4)\rangle = 
\left(\frac{x_{14}^2}{x_{24}^2}\right)^{a} \left(\frac{x_{14}^2}{x_{13}^2}\right)^{b} \frac{F(\ru,\rv)}{(x_{12}^2)^{\frac{1}{2} (\Del_1+\Del_2)} (x_{34}^2)^{\frac{1}{2} (\Del_3+\Del_4)}}, 
\ee
where $x_{ij} \equiv x_i-x_j$, $\Del_{ij} \equiv \Del_i-\Del_j$, $a \equiv -\frac{\Del_{12}}{2}$, $b \equiv \frac{\Del_{34}}{2},$ and the two independent conformally invariant cross ratios are:
\be
\ru = \frac{x_{12}^2 x_{34}^2}{x_{13}^2 x_{24}^2} = z \bz, \quad \rv =  \frac{x_{14}^2 x_{23}^2}{x_{13}^2 x_{24}^2} = (1-z)(1-\bz). 
\ee
Here we also introduced new variables $(z, \bz)$ that are more convenient for much of the work to follow; the reality of $(\ru,\rv)$ demands $z$ and $\bz$ are complex conjuates.
We can decompose \eqref{4pt-function-Para} into conformal partial waves:
\be\label{Def:CPW}
\langle \phi_1(x_1)\phi_2(x_2)\phi_3(x_3)\phi_4(x_4)\rangle =\sum_{\cO} \lam_{12\cO}\lam_{34\cO} W_\cO(x_1, x_2, x_3,x_4),
\ee
where $\lam_{12\cO}$ and $\lam_{34\cO}$ are the OPE coefficients, for the exchanged operator $\cO_{\Del, \ell}$ 
with scaling dimension $\Del$ and space-time spin $\ell$. The associated conformal block $G_{\cO}(\ru, \rv)$ 
is related to conformal partial wave by
\be\label{Def:ConfBlock0}
W_\cO(x_1,x_2,x_3,x_4) = \left(\frac{x_{14}^2}{x_{24}^2}\right)^{a} \left(\frac{x_{14}^2}{x_{13}^2}\right)^{b} \frac{G_\cO(\ru,\rv)}{(x_{12}^2)^{\frac{1}{2} (\Del_1+\Del_2)} (x_{34}^2)^{\frac{1}{2} (\Del_3+\Del_4)}},
\ee
hence $F(\ru,\rv) = \sum_{\cO}\lam_{12\cO}\lam_{34\cO}G_\cO(\ru,\rv)$.
We will also express $G_{\mathcal{O}}(\ru, \rv) = G_{\Del,\ell}(\ru,\rv) $ in terms of  $(z,\bz)$ as
\be\label{Def:ConfBlock1}
G_{\Del, \ell}(\ru, \rv) = \Fpm(z,\bz) = \Fpm(\bz,z), \quad \veps =\frac{(d-2)}{2},  \quad \lam_\pm = \frac{(\Del \pm l)}{2},
\ee
which are manifestly symmetric under $z \leftrightarrow \bz$ exchange, and where $\lambda_\pm$ are continuous parameters controlling  the asymptotic behavior of $\Fpm(z,\bz)$.
\paragraph{}
In this note we will study the actions of quadratic and quartic conformal Casimir operators on the conformal blocks involving scalar operators, chiral primary scalar operators, and primary operators with space-time spins.
To this end we define the following differential operators \cite{Dolan:2003hv}:
\bea
&&D_z(a, b, c) = z^2 (1-z) \frac{\partial^2}{\partial z^2} - ((a+b+1)z^2-c z)\frac{\partial}{\partial z} - ab z, \label{Def:Dzab} \\
&&D_\bz(a, b, c) = \bz^2 (1-\bz) \frac{\partial^2}{\partial \bz^2} - ((a+b+1)\bz^2-c z)\frac{\partial}{\partial \bz} - ab \bz. \label{Def:Dbzab}
\eea
Notice that the eigenvalue equation for $D_z(a, b, c)$ is given by:
\be\label{eigenDz}
D_z(a, b, c) g_\lam(a, b, c, z) = \lam(\lam+c-1)   g_\uprho(a, b, c, z) \Longrightarrow g_\lam(a, b,c, z) = z^\lam {}_2F_1(a+\lam, b+\lam, c+2\lam, z),
\ee
where ${}_2 F_1(a,b,c,z)$ is the hypergeometric function and the eigenvalue $\lam$ determines the $z \to 0$ asymptotic behavior of $g_\lam(a,b,c,z)$. 
We also have an identical eigenvalue equation for $D_\bz(a,b,c)$ operator. 
Using \eqref{Def:Dzab} and \eqref{Def:Dbzab}, we can define the main second order mixed partial differential operators that will be used subsequently for defining the action of various quadratic Casimir operators:
\be
\Del^{(\veps)}_2 (a, b, c) = D_z(a, b, c)+ D_\bz(a, b, c) + 2\veps\frac{z\bz}{z-\bz} \left((1-z)\frac{\partial}{\partial z}-(1-\bz)\frac{\partial}{\partial \bz}\right).\label{Def:Delabc}
\ee
The action of the quadratic Casimir operator $\hat{C}_2=\frac{1}{2} J_{AB} J^{AB}$ on a scalar primary conformal block \eqref{Def:ConfBlock1} can be written in terms of following eigenvalue equation \cite{Dolan:2003hv}:
\bea\label{Def:quadraticCasimir}
&&\Del^{(\veps)}_2 (a, b, 0) \Fpm(z,\bz) = c_2(\lam_+,\lam_-) \Fpm(z,\bz),  \nonumber  \\
&& c_2(\lam_+,\lam_-) = \lam_+(\lam_+-1)+\lam_-(\lam_- -1-2\veps).
\eea
We refer readers to Appendix \ref{App:Embedding} for an explanation of the Lorentz generators $J_{AB}$ and the embedding space formalism, 
and perhaps the most straightforward manner for deriving this differential equation.
\paragraph{}
Moreover, we can also define the quartic conformal Casimir operator $\hat{C}_4=\frac{1}{2} J_{AB}J^{BC}J_{CD}J^{DA}$. 
Its action on the scalar conformal block can be computed readily in the embedding coordinates and leads to the following eigenvalue equation \cite{Dolan:2011dv}:
\bea\label{Def:quarticCasimir}
&&\Del_4^{(\veps)}(a, b,0) \Fpm(z,\bz) = c_4(\lam_+,\lam_-) \Fpm(z,\bz),\\
&& c_4(\lam_+, \lam_-) =(\lam_+-\lam_-)(\lam_+-\lam_- +2\veps)(\lam_+ +\lam_- - 1) (\lam_+ +\lam_- - 1-2\veps),
\eea
where
\be\label{Def:Del4abc}
\Del_4^{(\veps)}(a, b,c) =\left(\frac{z\bz}{z-\bz}\right)^{2\veps}[D_z(a,b, c)-D_\bz(a,b,c)] \left(\frac{z-\bz}{z\bz}\right)^{2\veps}[D_z(a,b,c)-D_\bz(a,b, c)].
\ee
By the definition of Casimir operators, $\Del_2^{(\veps)}(a, b,0)$ and $\Delta_4^{(\veps)}(a,b,0)$ must commute and the scalar conformal block $F_{\lam_+,\lam_-}^{(\veps)}(z,\bz)$ is their mutual eigenfunction. More generally, we can have
\be\label{Commutation1}
\left[\Del_2^{(\veps)}(a,b, c), \Del_4^{(\veps)}(a, b, c)\right]=0. 
\ee
Notice that in the limit $\veps=0$, i. e. $d=2$, both $\Del_2^{(0)}(a,b,c)$ and $\Del_4^{(0)}(a,b,c)$ simplify into sum and product of purely $z$ and $\bz$ dependent ordinary differential operators,
we can build their common eigenfunctions using products of the solution to eigenvalue equation \eqref{eigenDz}.  
The scalar conformal blocks in this case are built from hypergeometric functions ${}_2F_1(a,b,c,z)$.
\paragraph{}
The main insight of \cite{Isachenkov:2016gim} was to realize that we can relate the conformal block $\Fpm(z, \bz)$ with the eigenfunction $\psipm(u, \bu)$ of certain quantum integrable Hamiltonian through an appropriate  ``gauge transformation''. Here we can be slightly more general and consider the following transformation of the operator $\Del^{(\veps)}_2(a,b,c)$ defined in \eqref{Def:Delabc}:
\be\label{TransDelabc}
\Del_2^{(\veps)}(a,b,c) \longrightarrow \chi^{(\veps)}_{a,b,c}(z,\bz) \Del_2^{(\veps)}(a,b,c) \frac{1}{\chi^{(\veps)}_{a,b,c}(z,\bz)}
\ee
and perform the change of variables:
\be
z(u) = -\frac{1}{\sinh^2 {u}}, \quad \bz(\bu) = -\frac{1}{\sinh^2{\bu}},\label{CoordTrans1}
\ee
where $z(u)$ and $\bz(u)$ are double cover maps invariant under $(z,\bz)\leftrightarrow(-z,-\bz)$\footnote{It is interesting to note that if we make the identification $\rho^{1/2}= e^{u+i\frac{\pi}{2}}$, $\bar{\rho}^{1/2} = e^{\bu-i\frac{\pi}{2}}$, 
we recover the ``radial coordinates'' used in \cite{Hogervorst:2013}.}. The specific choice of gauge transformation function $\chi^{(\veps)}_{a,b,c}$ is chosen to be:  
\bea
\chi^{(\veps)}_{a,b,c}(z(u),\bz(u)) &=& \frac{[(1-z(u))(1-\bz(\bu))]^{\frac{a+b-c}{2}+\frac{1}{4}}}{[z(u)\bz(\bu)]^{\frac{1-c}{2}}}  \left[\frac{|z(u)-\bz(\bu)|}{z(u)\bz(\bu)}\right]^\veps,\nn\\
&=&\frac{[\cosh u\cosh\bu]^{a+b-c+\frac{1}{2}}}{[\sinh u\sinh \bu]^{a+b-\frac{1}{2}}}[\sinh(u+\bu)\sinh(u-\bu)]^\veps.
\label{GaugeTrans1}
\eea
Now if we parameterize the eigenvalue equation for the quadratic Casimir operator $\Del^{(\veps)}_2(a, b, c)$ as:
\be\label{eigenDelabc}
\Del^{(\veps)}_2(a,b,c) F^{(\veps)}_{\lam_+,\lam_-}(z,\bz)= \frac{1}{2}\cE_{a,b,c}(\lam_+, \lam_-, \eps) F^{(\veps)}_{\lam_+,\lam_-}(z,\bz),
\ee
where the precise form of the eigenvalue $\frac{1}{2}\cE_{a,b,c}(\lam_+, \lam_-, \eps)$ is theory dependent
and it will control the $z \to 0, \bz\to 0$ asymptotic of the eigenfunction with $z\to 0$ taken first. 
More explicitly, if:
\be
\lim_{z\to 0, \bz\to 0} F_{\lam_+,\lam_-}(z,\bz)\sim z^{\lam_+} \bz^{\lam_-} 
\rightarrow
\frac{1}{2} \cE_{a,b,c}(\lam_+, \lam_-, \eps) = \lam_+(\lam_+ + c-1)+\lam_-(\lam_-+c-1-2\veps). \;\;\;
\ee   
Following the transformation \eqref{TransDelabc}, we need to correspondingly transform $F^{(\veps)}_{\lam_+, \lam_-}(z,\bz)$ as\footnote{Here we again suppress the $(a,b,c)$ dependence on $\psi^{(\veps)}_{\lam_+,\lam_-}(u,\bu)$ to avoid cluttered notation; however from \eqref{eigenDz} we clearly observe their dependences.}: 
\be
\psipm(u, \bu)= \chi^{(\veps)}_{a,b,c}(z(u),\bz(u))F_{\lam_+,\lam_-}^{(\veps)}(z(u), \bz(\bu)).
\label{gaugeTrans1}
\ee
This gauge transformation allows us to map the eigenvalue equation \eqref{eigenDelabc} into the following partial differential equation:
\bea\label{Hamiltonian1}
&&\frac{1}{4}\left(\frac{\partial^2}{\partial u^2}+\frac{\partial^2}{\partial \bu^2}\right) \psipm(u,\bu)
-\frac{\veps(\veps-1)}{2}  \left(\frac{1}{\sinh^2{(u-\bu)}}+ \frac{1}{\sinh^2{(u+\bu)}}   \right) \psipm(u,\bu)\nn\\
&&- \frac{1}{4}\left(\frac{((a-b)^2-\frac{1}{4})}{\sinh^2 u}-\frac{((a+b-c)^2-\frac{1}{4})}{\cosh^2 u}+  \frac{((a-b)^2-\frac{1}{4})}{\sinh^2 \bu}-\frac{ ((a+b-c)^2-\frac{1}{4})   }{\cosh^2\bu} \right) \psipm(u,\bu) \nn\\
&& =\left(\frac{1}{2}(1-c)^2+\veps(\veps+1-c)+\frac{1}{2}\cE_{a,b,c}(\lam_+,\lam_-,\veps)\right)\psipm(u,\bu)\eea
which can be verified directly using chain-rule differentiation.
The special case of scalar conformal block corresponds to setting $c=0$ and $\cE_{a,b,c}(\lam_+,\lam_-,\veps) =2 c_2(\lam_+,\lam_u)$.
The equation \eqref{Hamiltonian1} was recognized as the time independent Schr{\"o}dinger equation of  
hyperbolic Calogero-Sutherland system for reflection group ${\rm BC}_2$ \cite{Isachenkov:2016gim}, which we will discuss in more details momentarily. 
We will prove explicitly that the commuting quartic Casimir $\Del_4^{(\veps)}(a,b,c)$ can also be expressed in terms of the commuting higher conserved charges of the same integrable system, 
firmly established the connections with the quantum integrable systems.
When $\veps=0$, the $(u\pm\bu)$ dependent mixed potential terms vanish, 
the eigenvalue equation reduces to those of so-called P\"oschl-Teller potential, such that the solutions can be explicitly given by hypergeometric functions as we have seen in \eqref{eigenDz}. 
Moreover when $\veps=1$, the interacting potential terms between $u$ and $\bu$ also vanish, 
this implies in terms of $(u,\bu)$ coordinates, the corresponding wave functions for 2d and 4d scalar conformal blocks are both factorizable and solved by P\"oschl-Teller Hamiltonian.


\section{Dunkl Operators, Conserved Charges and Casimir Operators}
\paragraph{}
Here we follow \cite{Finkel:2002xf} to explicitly introduce the Dunkl operators and commuting conserved charges for the  hyperbolic Calogero-Sutherland (CS) spin chain with ${\rm BC}_2$ symmetry containing two spin degrees of freedom encoded in the two independent coordinates $(u, \bu)$.
Let us begin by defining the following reflection and permutation operators acting on an arbitrary two variable function $f(u, \bu)$:
\bea
&&\rK_u f(u,\bu) = f(-u, \bu), \quad  \rK_\bu f(u, \bu) = f(u,-\bu),\nn\\
&&\rK_{u\bu}f(u,\bu) = f(\bu, u), \quad \tilde{\rK}_{u\bu} f(u,\bu) = \rK_u\rK_\bu\rK_{u\bu} f(u,\bu) = f(-\bu,-u),
\eea
such that $\rK_{u\bu} = \rK_{\bu u}$ and $\trK_{u\bu}=\trK_{\bu u}$.
The Dunkl operators of this system are given by:
\bea \label{Def:DunklOpu}
\hJua=   \frac{\partial}{\partial u} - [{\bf{b} }(1+\coth u)+\bb'(1+\tanh u)]\rK_{u}  \;\;\;\;\;\;\;\;  \;\;\;\;\;\;\;\;\;\;\;\;\;\;\;\;\;\;\;\;\;\;\;\;  \\
\;\;\;\;\;\;\;\;\;\;\;\;\;\;\;\; -\ba[(1+\coth(u+\bu))\tilde{\rK}_{u\bu}+ (1+\coth(u-\bu))\rK_{u\bu}],  \nonumber
\eea
\bea \label{Def:DunklOpbu}
\hJbua =  \frac{\partial}{\partial \bu} - [\bb(1+\coth \bu)+\bb'(1+\tanh \bu)]\rK_{\bu} \;\;\;\;\;\;\;\; \;\;\;\;\;\;\;\;\;\;\;\;\;\;\;\;\;\;\;\;\;\;\;\; \\
\;\;\;\;\;\;\;\;\;\;\;\;\;\;\;\; -\ba[(1+\coth(u+\bu))\tilde{\rK}_{u\bu} -(1+\coth(u-\bu))\rK_{u\bu} ], \nonumber
\eea
where $\ba, \bb$, and $\bb'$ are real parameters. The Dunkl operators commute:
\be
\left[\hJua, \hJbua\right] = 0.
\ee
From these we can construct the following algebraically independent commuting integrals of motion:
\be\label{Def:I2p}
\hcI_{2p} = -\left(\hJua\right)^{2p}-\left(\hJbua\right)^{2p}, \quad p=1, 2
\ee
such that:
\be\label{Commutation2}
\left[\hcI_2, \hcI_4\right] = 0.
\ee
This is the quantum mechanical analogue of the Liouville integrability condition. 
In fact, for general $p \in {\mathbb N}$ we can also construct higher commuting conserved charges $\hcI_{2p}$, $p\ge 3$; for this two degrees-of-freedom system, however, they are no longer algebraically independent and can be expressed in terms of $\hcI_2$ and $\hcI_4$. 
In particular, the Hamiltonian of hyperbolic CS spin chain of ${\rm BC}_2$ is given by:
\bea
H_{{\rm BC}_2} =  \hcI_2 &=& -\left(\frac{\partial^2}{\partial u^2} + \frac{\partial^2}{\partial \bu^2}\right )+\left(\frac{\bb(\bb-\rK_u)}{\sinh^2 u}-\frac{\bb'(\bb'-\rK_u)}{\cosh^2 u}\right) + \left(\frac{\bb(\bb-\rK_\bu)}{\sinh^2 \bu}-\frac{\bb'(\bb'-\rK_\bu)}{\cosh^2 \bu}\right)\nn\\
&+& 2\ba \left(\frac{(\ba-\rK_{u\bu})}{\sinh^2(u-\bu)}+  \frac{(\ba-\trK_{u\bu})}{\sinh^2(u+\bu)} \right).
\eea
Now we consider $H_{{\rm BC}_2}=\hcI_2$ acting on the wave function $\psipm(u, \bu)$ defined in \eqref{gaugeTrans1}.  Both $\chi^{(\veps)}_{a,b,c}(z(u),\bz(\bu))$ and $\Fpm(z(u),\bz(u))$ are manifestly invariant under the actions of the reflection and permutation operators due to   \eqref{Def:ConfBlock1} and   \eqref{CoordTrans1}, and thus
\begin{gather}\label{SymmetryConditions}
\rK_u \psipm(u, \bu)= \rK_\bu\psipm(u,\bu) = \rK_{u\bu}\psipm(u,\bu)  = \trK_{u\bu}\psipm(u,\bu) = \psipm(u,\bu).
\end{gather}
If we now make the following identification of parameters:
\be
\ba = \veps, \quad \bb = (a-b)+\frac{1}{2}, \quad \bb' = (a+b-c)+\frac{1}{2},
\ee
the eigenvalue equation for $\Del^{(\veps)}_2(a,b,c)$ \eqref{eigenDelabc} 
is now mapped to the Schr{\"o}dinger equation for $\hcI_2$ as: 
\bea\label{Relation1}
\chi^{(\veps)}_{a,b,c}(z,\bz)\Del_2^{(\veps)}(a,b,c)F_{a,b,c}^{(\veps)}(z,\bz) &=& -\left(\frac{1}{4} \cI_2 +\frac{1}{2}(1-c)^2+\veps(\veps+1-c) \right) \psipm(u,\bu)   \nn\\
&=&  
\frac{1}{2} \cE_{a,b,c}(\lam_+, \lam_-, \eps)  \psipm(u,\bu).
 \eea
In other words, $\psipm(u,\bu)$ now becomes an eigenfunction of $H_{{\rm BC}_2}=\cI_2$ with eigenvalue:
\be
-2(1-c)^2-4\veps(\eps+1-c)-2\cE_{a,b,c}(\lam_+, \lam_-, \eps).
\ee
Up to an overall constant shift in the eigenvalue, we can also express the mapping \eqref{Relation1} in the operator form as:
\be\label{TransformedDel2}
\chi^{(\veps)}_{a,b,c}(z,\bz)\Del_2^{(\veps)}(a,b,c) \frac{1}{\chi^{(\veps)}_{a,b,c}(z,\bz)} = -\frac{1}{4}  \cI_2.
\ee
Notice that the un-hatted operators indicate we have replaced all $\rK_u,\rK_\bu, \rK_{u\bu}$ and $\trK_{u\bu}$ in the corresponding hatted ones by the identity operators,
which is valid when we consider their trivial actions on a symmetric function $f(u,\bu)$. Upon setting $c=0$ and $\cE_{a,b,c}(\lam_+, \lam_-, \eps) = 2c_2(\lam_+,\lam_-)$, we thus recover the results in \cite{Isachenkov:2016gim}.
We now wish to prove that, like the action of $\Del_2^{(\veps)}(a,b,c)$ on $F_{a,b,c}^{(\veps)}(z,\bz)$, the action of  $\Del_4^{(\veps)}(a,b,c)$ on $F_{a,b,c}^{(\veps)}(z,\bz)$ can be mapped to the combination of commuting $\cI_2$ and $\cI_4$.
\paragraph{}
As a warm up, let us consider $\veps=\ba=0$ limit, which dictates us to relate $\hJu0$ and $\hJbu0$ with the differential operators $D_z(a, b, c)$ and $D_\bz(a, b, c)$ defined in \eqref{Def:Dzab}
and \eqref{Def:Dbzab}. Explicitly, consider the following equations:
\bea
\left(J_u^{(0)}\right)^2 \psi^{(0)}_{\lam_+\lam_-}(u,\bu)= \left(\frac{\partial^2}{\partial u^2} - \frac{\bb(\bb-1)}{\sinh^2 u}+\frac{\bb'(\bb'-1)}{\cosh^2 u}\right) \psi^{(0)}_{\lam_+\lam_-}(u,\bu)  , \\
\left(J_\bu^{(0)}\right)^2 \psi^{(0)}_{\lam_+\lam_-}(u,\bu)= \left(\frac{\partial^2}{\partial \bu^2} - \frac{\bb(\bb-1)}{\sinh^2 \bu}+\frac{\bb'(\bb'-1)}{\cosh^2 \bu}\right) \psi^{(0)}_{\lam_+\lam_-}(u,\bu),
\eea
where we have used the fact $ \psi^{(0)}_{\lam_+\lam_-}(u,\bu)$ is invariant under the reflection operators $\rK_u$ and $\rK_\bu$ (while in the intermediate step it is necessary to keep track of them to cancel the first derivative terms).
If we consider their corresponding actions in $(z, \bz)$ coordinates using gauge transformation $\chi^{(0)}_{a,b,c}(z,\bz)$, we can obtain the following relations:
\begin{eqnarray}\label{DzDbz-relation}
&&\frac{1}{4} \left(\Ju0\right)^2 -\frac{1}{4}(1-c)^2   = \chi^{(0)}_{a,b,c}(z, \bz) D_z(a,b,c)\frac{1}{\chi^{(0)}_{a,b,c}(z, \bz)}, \\
&&\frac{1}{4} \left(\Jbu0\right)^2- \frac{1}{4}(1-c)^2  = \chi^{(0)}_{a,b,c}(z, \bz) D_\bz(a,b,c)\frac{1}{\chi^{(0)}_{a,b,c}(z, \bz)}, 
\end{eqnarray}
valid for acting on any $\rK_u$ and $\rK_\bu$ invariant $f(u,\bu)$.
Now for generic $\veps$, we can sandwich the quartic Casimir defined in \eqref{Def:quarticCasimir} inbetween $\chi^{(\veps)}_{a,b,c}(z,\bz)$ and $\frac{1}{\chi^{(\veps)}_{a,b,c}(z,\bz)}$, 
using \eqref{DzDbz-relation} we have:
\bea\label{TransformedDel4}
&&\chi^{(\veps)}_{a,b,c}(z,\bz)\Del_4^{(\veps)}(a,b,c)\frac{1}{ \chi^{(\veps)}_{a,b,c}(z,\bz)}\nn\\
&& =  \frac{\chi^{(0)}_{a,b,c}(z,\bz)}{\tveps(z,\bz)}[D_z(a,b,c)-D_\bz(a,b,c)][\tveps(z,\bz)]^2 [D_z(a,b,c)-D_\bz(a,b,c)]\frac{1}{\chi^{(0)}_{a,b,c}(z,\bz)}\frac{1}{ \tveps(z,\bz)}, \nn\\
&&= \frac{1}{16}\left( \frac{1}{\tveps(z,\bz)}\left[\left(\Ju0\right)^2-  \left(\Jbu0\right)^2 \right]\tveps(z,\bz)\right) 
\left(\tveps(z,\bz) \left[\left(\Ju0\right)^2-  \left(\Jbu0\right)^2 \right] \frac{1}{\tveps(z,\bz)}\right),\nn\\
\eea
where we have introduced:
\be\label{Def:teps}
t^{(\veps)}(z,\bz) = \frac{\chi^{(\veps)}_{a,b,c}(z,\bz)}{ \chi^{(0)}_{a,b,c}(z,\bz)} = \left[\frac{|z(u)-\bz(\bu)|}{z(u)\bz(\bu)}\right]^\veps.
\ee
Clearly from \eqref{Commutation1}, the gauge-transformed $\Del_2^{(\veps)}(a,b,c)$ and $\Del_4^{(\veps)}(a,b,c)$ in the LHS of \eqref{TransformedDel2} and \eqref{TransformedDel4} 
remain commuting, and RHS of \eqref{TransformedDel2} is given by $\cI_2$, the remaining task is to express RHS of \eqref{TransformedDel4} 
or more precisely its action on a symmetric function $\psipm(u,\bu)$ in terms of $\cI_2$ and $\cI_4$.
\paragraph{}
Let us define the following combinations:
\bea
&&\hLup = \hJueps+\hJbueps + 2\veps \trK_{u\bu} = \hJu0+\hJbu0-2\veps\coth(u+\bu)\trK_{u\bu}, \label{Def:Lplus}\\
&&\hLum= \hJueps-\hJbueps + 2\veps \rK_{u\bu} = \hJu0-\hJbu0-2\veps\coth(u-\bu)\rK_{u\bu} \label{Def:Lminus},
\eea
which satisfy the non-trivial commutation relation:
\be\label{LpLm-commutator}
[\hLup,\hLum] = 2(\bb+\bb')(\rK_u-\rK_\bu) \trK_{u\bu}.
\ee
Now consider their combined action on the symmetric eigenfunction $\psipm(u,\bu)$:
\bea\label{Product1}
\frac{1}{4}{\hLum}{\hLup}\psipm(u,\bu)
&=&\frac{1}{4} \left[\left(\Ju0\right)^2-  \left(\Jbu0\right)^2 \right] \psipm(u,\bu)+\veps^2\coth(u-\bu)\coth(u+\bu)\psipm(u,\bu)\nn\\
&-&\frac{\veps}{2}\left[\coth(u-\bu)(\partial_u+\partial_\bu)+\coth(u+\bu)(\partial_u-\partial_\bu)\right] \psipm(u,\bu)\nn\\
&=& \frac{1}{4} t^{(\veps)}(z,\bz)  \left[\left(\Ju0\right)^2-  \left(\Jbu0\right)^2 \right] \frac{1}{t^{(\veps)}(z,\bz)}  \psipm(u,\bu)
\eea 
where in the first two lines we have used the invariant properties of $\psipm(u,\bu)$ in \eqref{SymmetryConditions}, 
while the last line can be verified through direct differentiation.
Acting ${\hLum}{\hLup}\psipm(u,\bu)$ with reflection and permutation operators:
\bea\label{SymmetryConditions2}
\rK_u{\hLum}{\hLup}\psipm(u,\bu) =+{\hLum}{\hLup}\psipm(u,\bu), \nn\\
\rK_\bu{\hLum}{\hLup}\psipm(u,\bu) =+{\hLum}{\hLup}\psipm(u,\bu),\nn\\
 \rK_{u\bu} {\hLum}{\hLup}\psipm(u,\bu) = -{\hLum}{\hLup}\psipm(u,\bu), \nn\\
 \trK_{u\bu}{\hLum}{\hLup}\psipm(u,\bu) = -{\hLum}{\hLup}\psipm(u,\bu).
\eea
This implies:
\bea\label{Product2}
&&\frac{1}{16}\hLum\hLup\hLum\hLup \psipm(u,\bu)\nn\\
&&=\frac{1}{4} \left[\left(\Ju0\right)^2-  \left(\Jbu0\right)^2 \right]  \frac{1}{4}{\hLum}{\hLup}\psipm(u,\bu) + {\veps^2}\coth(u-\bu)\coth(u+\bu)\frac{1}{4}{\hLum}{\hLup}\psipm(u,\bu)\nn\\
&&+\frac{\veps}{2}\left[\coth(u-\bu)(\partial_u+\partial_\bu)+\coth(u+\bu)(\partial_u-\partial_\bu)\right]\frac{1}{4}{\hLum}{\hLup}\psipm(u,\bu)\nn\\
&&= \frac{1}{4} \frac{1}{t^{(\veps)}(z,\bz)}  \left[\left(\Ju0\right)^2-  \left(\Jbu0\right)^2 \right] \frac{1}{4}{t^{(\veps)}(z,\bz)} {\hLum}{\hLup}\psipm(u,\bu)\nn\\
&&= \frac{1}{16} \left( \frac{1}{\tveps(z,\bz)}\left[\left(\Ju0\right)^2-  \left(\Jbu0\right)^2 \right]\tveps(z,\bz)\right) 
\left(\tveps(z,\bz) \left[\left(\Ju0\right)^2-  \left(\Jbu0\right)^2 \right] \frac{1}{\tveps(z,\bz)}\right)\psipm(u,\bu),\nn\\
\eea
where in the third line above we have used the properties in \eqref{SymmetryConditions2} to flip the crucial sign for the first derivative comparing with \eqref{Product1}, giving the opposite power dependence of $t^{(\veps)}(z,\bz)$ in the fourth line. 
This also makes clear that the role of permutation operators $\rK_{u\bu}$ and $\trK_{u\bu}$ in establishing such a correspondence between commuting Casimir operators and conserved charges.
\paragraph{}
We have just demonstrated that the action of gauge transformed $\Del_4^{(\veps)}(a,b,c)$ in \eqref{TransformedDel4} can be expressed as simple product of $\hLup$ and $\hLum$. Be expanding the product, we obtain:
\bea\label{Explicit-Product2}
&&\frac{1}{16}\hLum\hLup\hLum\hLup \psipm(u,\bu) \nn\\
&&= \left(\frac{1}{16}\left[\left(\Jueps\right)^2 -  \left(\Jbueps\right)^2  \right]^2 - \frac{\veps^2}{2} \left[\left(\Jueps\right)^2 +  \left(\Jbueps\right)^2  \right]+\veps^4\right) \psipm(u,\bu) \nn\\
&&=\left(\frac{1}{8} \left[\left(\Jueps\right)^4 +  \left(\Jbueps\right)^4  \right] -\frac{1}{16} \left[\left(\Jueps\right)^2 + \left(\Jbueps\right)^2  \right]^2-\frac{\veps^2}{2}  \left[\left(\Jueps\right)^2 + \left(\Jbueps\right)^2  \right]+\veps^4\right)  \psipm(u,\bu)\nn\\
&&= \left(-\frac{1}{8} \cI_4 +\frac{1}{16}\cI_2^2+\frac{\veps^2}{2}\cI_2+\veps^4\right)\psipm(u,\bu).
\eea
In operator form, we have explicitly:
\be\label{TransformedDel4-final}
\chi^{(\veps)}_{a,b,c}(z,\bz)\Del_4^{(\veps)}(a,b,c)\frac{1}{ \chi^{(\veps)}_{a,b,c}(z,\bz)} =  -\frac{1}{8} \cI_4 +\frac{1}{16}\cI_2^2+\frac{\veps^2}{2}\cI_2+\veps^4,
\ee
comparing with \eqref{TransformedDel2}, we can clearly see that commuting Casimirs \eqref{Commutation1} acting on the conformal blocks for scalar primary in d-dimensional conformal field theories implies independent commuting conserved charges \eqref{Commutation2} in the quantum integrable system, and vice versa. 
Our calculation implies that if the action of the quadratic Casimir operator on correlation function containing a certain class of external primary fields can be cast into an eigenvalue problem as in \eqref{Def:quadraticCasimir}, 
both commuting quadratic and quartic Casimir operators can be mapped to the commuting integrals of motion of certain quantum integrable systems.

\section{Generalizations to SCFTs with Four and Eight Supercharges}
\paragraph{}
Here we would like to generalize our earlier results to the conformal blocks in $d$-dimensional superconformal field theories with four and eight supercharges. Selected previous work on superconformal blocks in four dimensions includes \cite{SCFT refs}. 
\paragraph{}
We begin by considering the case of four supercharges, corresponding to $\mathcal{N}=1$ in four dimensions. Specifically, we consider the four point correlation function of the lowest scalar component in a superconformal scalar primary $\Phi_i(x_i),~i=1,2,3,4,$ of scaling dimension $\Delta_i$ and R-charge $q_i$. We will also impose the additional constraint that $\Phi_{1,3}(x)$ are chiral, i. e. $Q_\alpha^+ \Phi_{1,3} =0$, such that $\Del_{1,3} = \frac{(d-1)}{2} q_{1,3}$.
In the non-supersymmetric case, we used the embedding space formalism to write down a differential equation for conformal blocks. 
In the current case, we make use of a supersymmetric generalization of this idea: the superembedding space. 
Please refer to Appendix \ref{App:Embedding} again for an overview and detailed references.
The action of the quadratic superconformal Casimir on this class of superconformal blocks has been worked out in \cite{Bobev:2015jxa},
which can again be expressed as the following eigenvalue equation: 
\bea
&&\Del_2^{(\veps)}(a+1, b, 1) \cF_{\lam_+\lam_-}^{\cN=1}(z, \bz)= c_2^{\cN=1}(\lam_+,\lam_-)\cF_{\lam_+\lam-}^{\cN=1}(z, \bz),\nn\\
&&c_2^{\cN=1}(\lam_+,\lam_-) =  \lam_+^2+\lam_-(\lam_- -2\veps).\label{Def:Casimir2-4}  
\eea
The superscript ``$\cN=1$'' is to be consistent with four-dimensional counting of the number of supersymmetries, however it should be understood that $\veps$ enters the above equation as a parameter hence this expression applies to general d-dimensional SCFTs permitting four supercharges. Moreover, the R-charge dependences cancel out on the both sides of superconformal Casimir equation \eqref{Def:Casimir2-4}.
\paragraph{}
Using the gauge transformation \eqref{GaugeTrans1}, we can easily see that identical Calogero-Sutherland integrable system arises in this class of chiral primary conformal block, 
with only simple shifts in the coefficients for the $1/\sinh^2 u$ and $1/\sinh^2 \bu$ potential terms and the eigenvalue comparing with scalar primary case.
This shift in the coefficients $(a, b)$ is perfectly consistent with the multiplicative relation between the superconformal and conformal blocks proposed in \cite{Bobev:2015jxa}:
\be \label{superblockstoblocksN=1}
\mathcal{G}^{\cN=1}_{\Delta, \ell}(\ru,\rv) = u^{-\frac{1}{2}} G^{\Delta_{12}-1, \Delta_{34}-1}_{\Delta+1,\ell}(\ru,\rv).
\ee
Here we have explicitly indicated the shift in the scaling dimensions of the external and internal operators in the conformal block on RHS.
We can also view the additional factor $u^{-\frac{1}{2}} =(z\bz)^{-\frac{1}{2}}$ as part of the gauge transformation, 
it is natural therefore to expect we can relate them to the same class of quantum integrable systems. 
By the connection with the quantum integrable systems, we can now for example conversely propose 
that the quartic Casimir acting on $\cFpm^{\cN=1}(z,\bz)$ is given by the following differential operator:
\be
\Del_4^{(\veps)}(a+1, b, 1)=\left(\frac{z\bz}{z-\bz}\right)^{2\veps}[D_z(a+1,b, 1)-D_\bz(a+1,b,1)] 
\left(\frac{z-\bz}{z\bz}\right)^{2\veps}[D_z(a+1,b,1)-D_\bz(a+1,b, 1)]. 
\ee
\paragraph{}
We now turn our attention to the theories with eight supercharges, where a nearly identical story happens. While in general one needs to extend the superembedding space for four supercharges to describe SCFTs with extended supersymmetry, as discussed in Appendix \ref{App:Embedding}, however our current formalism is sufficient for describing operators annihilated by all supersymmetries of one chirality\footnote{For $\mathcal{N} = 4$, this constraint is overly restrictive, and satisfied only by the identity.}. 
For $\mathcal{N}=1$, we needed to impose that $\Phi_{1,3}$ were chiral. For $\mathcal{N}=2$, superconformal invariance dictates that the only OPE decomposition channel that will have an associated superconformal block is for the OPE between a chiral and an anti-chiral field \cite{Lemos:2015awa}. 
We therefore consider four point correlation functions of the form (or more precisely correlators of the lowest scalar components):
\be\label{n2fourpoint}
\langle \Phi_1(P_1) \bar{\Phi}_2(P_2)\Phi_3(P_3)\bar{\Phi}_4(P_4)\rangle,
\ee
where the $R$-charges are restricted so that $q_4=q_1$ and $q_3=q_2$.
Now in $d=4$ dimensions or $\veps=1$, superembedding formalism allows us to derive a partial differential equation for this class of superconformal block,
and they become the eigenfunctions of quadratic Casimir operator $\Del^{(1)}_2(a, a+2, 2)$  \cite{Lemos:2015awa},  
we would like to conjecture this holds for other $\veps$ hence SCFTs in other dimensions permitting eight supercharges as well. 
The gauge transformation to the corresponding ${\rm BC}_2$ Calogero-Sutherland system carries through identically using \eqref{GaugeTrans1}. 
Once again, the superconformal block (related to the non-supersymmetric conformal block by gauge transformation $\sim u^{-1} =(z\bz)^{-1}$) is described by the same quantum mechanical integrable system, with simple shifts in the coefficients for the potential terms.
\paragraph{}
We conclude these discussions by extending our work to more general supersymmetric correlation functions. We have thus far considered superconformal block involving chiral or anti-chiral primary fields, thank to the fact that in these cases one can use recurrence relations to recast the superconformal block in terms of a single nonsupersymmetric scalar conformal block up to an overall multiplicative factor. There is a straightforward extension of our results to the most general 4d $\mathcal{N}=1$ four-point functions of scalars $\langle \Phi_1\Phi_2\Phi_3\Phi_4 \rangle$, where the $\Phi_i$ have independent scaling dimensions and R-charges  \cite{Li:2016chh}\footnote{These fields are still subject to the condition that their total R-charge vanishes in order to preserve the $U(1)_R$ symmetry.}. In this case, the superconformal block is given in terms of of a linear combination of scalar conformal blocks. Translating into the notations we will use in next section:
\be\label{General4dsuperblock}
\cG^{r,\bar{r}}_{\Delta,\ell}(z,\bz) = \sum_{m,n=0, 1} c_{m,n} F_{\lambda_++m,\lambda_-+n}^{-\frac{r}{2},+\frac{\bar{r}}{2},0}(z,\bz),
\ee
where $\cG^{r,\bar{r}}_{\Delta,\ell}(z,\bz)$ is the ``most general'' 4d $\cN=1$ scalar superconformal block with $r=\Delta_1-\Delta_2$ and $\bar{r}=\Delta_3-\Delta_4$.
The functions $F_{\lambda_++m, \lambda_-+n}^ {-\frac{r}{2},+\frac{\bar{r}}{2},0}(z,\bz)$ are non-supersymmetric scalar blocks, and we have given the explicit values of $a=-\frac{r}{2},b=+\frac{\bar{r}}{2},c=0$. The constant expansion coefficients $c_{m,n}$ 
are functions of parameters such as the scaling dimensions and spins of the exchanged operators as well as external scaling dimensions, R-charges etc. 
It is clear that the individual scalar conformal blocks in the expansion can be recast into the eigenfunction of hyperbolic CS Hamiltonian of ${\rm BC}_2$ type with appropriate coefficients using the gauge transformation \eqref{GaugeTrans1}, this analysis is very similar to the conformal blocks with external space-time spins to be discussed next.

\section{Some Comments on Generalization to Correlation with Spins}
\paragraph{}
More generally, we wish to extend our analysis to four point correlation functions containing external operators with non-trivial space-time spins.
After the tensor decomposition of Lorentz representations, the number of  conformal partial waves goes linearly with the number of independent tensor structures $N_4$, 
while the number of conformal blocks for each exchange channel goes like $N_4\times N_4$.
This is in significant contrast with the scalar correlation functions, where each exchange channel only contains single conformal block.
\paragraph{}
We can, however, simplify the problem systematically by first restricting the tensor structures of the exchanged operators in the OPEs. This in turn allows us to express the conformal partial wave for each independent tensor structure through 
combination of certain space-time differential operators acting on a {\it single}, much simpler conformal partial wave.
For example, if we restrict the exchange operators to have only symmetric traceless tensor structure \cite{Costa:2011dw} then their associated conformal partial waves for independent tensor structures can be expressed as differential operators acting on the scalar conformal partial wave in \eqref{Def:CPW}. 
In four dimensions, the exchange operators  $\cO_{l,\bar{l}}$ are labeled in terms of its scaling dimension $\Del$ and Lorentz spins $(l,\bar{l})$.
For a non-negative integer $p$ which is the minimal value of the spins $l_i+\bar{l}_i$ of the external operators,
we can consider the exchange operators $\cO_{l,\bar{l}}$ transforming under the general Lorentz representations, 
which can be bosonic or fermionic for $|l-\bar{l}| = 0, 2, 4, \dots, p$ or $|l-\bar{l}|=0, 1,3, \dots, p$ respectively \cite{Echeverri:2016dun}. 
Conformal partial waves for all independent tensor decompositions are now generated by differential operators acting on a single {\it seed} conformal partial wave
or its conjugate:
\bea
W_{\cO_{l,l+p}}^{\rm seed}(x_i) &=& \left(\frac{x_{14}^2}{x_{24}^2}\right)^{\frac{\tau_{21}}{2}} \left(\frac{x_{14}^2}{x_{13}^2}\right)^{\frac{\tau_{34}}{2}} \frac{ \sum_{e=0}^{p} G_e^{(p)}(\ru,\rv) \bI^{e}_{42} \bJ^{p-e}_{42, 31} }{(x_{12}^2)^{\frac{1}{2} (\tau_1+\tau_2)} (x_{34}^2)^{\frac{1}{2} (\tau_3+\tau_4)}},\\
\overline{W}_{\overline{\cO}_{l+p,l}}^{\rm seed}(x_i) &=& \left(\frac{x_{14}^2}{x_{24}^2}\right)^{\frac{\tau_{21}}{2}} \left(\frac{x_{14}^2}{x_{13}^2}\right)^{\frac{\tau_{34}}{2}} \frac{ \sum_{e=0}^{p} \overline{G}_e^{(p)}(\ru,\rv) \bI^{e}_{42} \bJ^{p-e}_{42, 31} }{(x_{12}^2)^{\frac{1}{2} (\tau_1+\tau_2)} (x_{34}^2)^{\frac{1}{2} (\tau_3+\tau_4)}},
\eea
where $\tau_i = \Del_i + \frac{l_i+\bar{l}_i}{2}$ and $\tau_{ij} = \tau_i-\tau_j$.
$\bI_{ij}$ and $\bJ_{ij, kl}$ form an independent basis constructed from polarization spinors and label the $p+1$ independent tensor structures, 
as the seed conformal partial waves themselves can be expressed in terms the four point correlation functions containing two scalar and two tensor fields of the form
$\langle\phi_1(x_1)F_{2}^{(p,0)}\phi_3(x_3) \bar{F}_4^{(0,p)}(x_4) \rangle$. More details can be found in \cite{Echeverri:2016dun} (see also \cite{Echeverri:2015rwa}, \cite{Elkhidir:2014woa});
here we focus on the action of quadratic Casimir $\hat{C}_2$ on $W_{\cO_{l,l+p}}^{\rm seed}(x_i)$ (or equivalently $\overline{W}^{\rm seed}_{\overline{\cO}_{l+p,l}}(x_i)$).
While the entire $W^{\rm seed}_{\cO_{l,l+p}}(x_i)$ remains an eigenfunction due to conformal invariance, the Casimir $\hat{C}_2$ acts non-trivially on the independent tensors $\bI_{42}^e \bJ^{p-e}_{42,31}$ 
and effectively permutes them. There are therefore $p+1$ resultant coupled partial differential equations for the {\it seed} conformal blocks $G_{e}^{(p)}(x_i)$, $e=0, 1,\dots, p$. 
They can be succinctly summarized in the following matrix eigenvalue equation:
\bea \label{SeedBlock-Casimir}
&&\hat{\mathbb {C}}^{(p)} \cdot \vec{G}(z,\bz) = \frac{1}{2} E_l^p \vec{G}(z,\bz), \quad E^p_l =\Del(\Del-4)+l^2+(2+p)(l+\frac{p}{2}), \\
&&\vec{G}(z,\bz) = \left(G_0^{(p)}(z,\bz),  G_1^{(p)}(z,\bz), \dots, G_p^{(p)}(z,\bz)\right)^{\rm t},\label{Def:vecG}\\
&& \hat{\mathbb{C}}_{e,e}^{(p)} =  \Del^{(\frac{p+2}{2})}_2(a_e, b_e, c_e)+\frac{1}{2}\epsilon^p_e, \enspace \hat{\mathbb{C}}_{e-1, e}^{(p)} = A_e^p z\bz  L(a_{e-1}),\enspace 
\hat{\mathbb{C}}_{e,e+1}^{(p)} = B_e L(b_{e+1}),\label{Def:C-components}
\eea
where $L(\mu)$ is a first order differential operator:
\be\label{Def:Lopt}
L(\mu) = -\frac{1}{z-\bz}\left(z(1-z)\frac{\partial}{\partial z} - \bz(1-\bz)\frac{\partial}{\partial \bz} \right)+\mu,
\ee
and all other entries of $(p+1)\times (p+1)$ matrix differential operator $\hat{\mathbb{C}}$ vanish.
The various constant parameters above are given by:  
\bea\label{Def:parameters}
&& \eps_e^p =\frac{3}{4}p^2-(1+2e)p+2e(2+e),~~ A_e^p=2(p-e+1),~~B_e=\frac{e+1}{2},\nn\\ 
&& a_e=\frac{\Del_2-\Del_1}{2}+\frac{p}{4},~ b_e = \frac{\Del_3-\Del_4}{2}-\frac{p}{4}+(p-e),~c_e =(p-e).
\eea
\paragraph{}
This system of PDEs has been solved \cite{Echeverri:2016dun}, and we re-express the solutions in the form:
\be\label{seedblocksolutions}
G_{e}^{(p)}(z,\bz) =  \left( \frac{z\bar{z}}{z-\bar{z}} \right)^{2p} \sum_{(m,n)\in Oct^{(p)}_{\,e}} c^e_{m,n} {F}^{a_e, b_e, c_e}_{\frac{\Delta+l+p/2}{2}+m, \frac{\Delta-l+p/2}{2}-(p+1)+n}(z,\bz).
\ee
In this equation, the coefficients $c^e_{m,n}$ satisfy certain recursion relationships and are functions of $\Delta,l,a$ and $b$ (but not $p$ or $e$). 
The sum generically runs over points in a particular octagon $Oct^{(p)}_{\, e}$ of $N_p^e$ points, where
\be
N_p^e = 2p(2p-e)+(1+e)(3p+1-e).
\ee
We have also made the dependence of eigenfunction  $F_{\rho_1, \rho_2}^{a_e, b_e, c_e}(z,\bz)$  of $\Delta_{2}^{(1)}(a_e, b_e, c_e)$ on $a_e,b_e,$ and $c_e$ explicit, and suppressed the $\veps =1 $ depdendence for our purposes here.
\paragraph{}
In order to interpret these spinning conformal blocks in the quantum integrable system, we first make some additional observations about the differential operator $\Del^{(\veps)}_2(a, b, c)$. It can be shown \cite{Echeverri:2016dun}:
\be \label{deltak}
\left( \frac{z\bar{z}}{z-\bar{z}} \right)^{-k}  \Del^{(\veps)}_2(a,b,c) \left( \frac{z\bar{z}}{z-\bar{z}} \right)^k =  \Del^{(\veps-k)}_2(a,b,c) + k(k-2\veps+c-1) -k(k-2\veps+1)\frac{z\bar{z}(z+\bar{z})-2z\bar{z}}{(z-\bar{z})^2}.
\ee
A particularly useful case of the identity above is for $k=\veps=1$, where the cross terms drop off:
\be
\Del^{(\veps)}_2(a,b,c)  \left( \frac{z\bar{z}}{z-\bar{z}} \right) F^{(1)}_{\rho_1,\rho_2}(z,\bz)= \left(\frac{1}{2}\cE_{a,b,c}(\rho_1, \rho_2, 1) +(c-2)\right)   \left( \frac{z\bar{z}}{z-\bar{z}} \right) F^{(1)}_{\rho_1,\rho_2}(z,\bz).
\ee
Moreover from 
\be
(z\bz)^{-k}   \Del^{(\veps)}_2(a, b, c) (z\bz)^k =  \Del^{(\veps)}_2(a+k, b+k, c+2k) + 2k\left(k-1+c-\veps\right),
\ee
we also deduce that
\be
(z\bz)^k F^{a+k,b+k,c+2k}_{\rho_1,\rho_2}(z,\bz) = F^{a,b,c}_{\rho_1+k,\rho_2+k}(z,\bz).
\ee
Consequently, we can always relate a function $F^{a, b, c}_{\rho_1,\rho_2}(z,\bz)$ with $c\neq0$ to a 4d scalar conformal block having vanishing $c$. We see that the seed conformal blocks \eqref{seedblocksolutions} that are multiplied with the various tensor structures are therefore given as an overall function times a linear combination of scalar conformal blocks, each one having a different eigenvalue under the quadratic conformal Casimir operator. The remaining steps are a straightforward exercise. Applying an appropriate gauge transformation \eqref{GaugeTrans1} (with $a,b,c\rightarrow a_e,b_e,c_e$) to both sides of \eqref{seedblocksolutions},
we find that the ``seed blocks'' in Hilbert space are given by an overall function times a linear combination of Calogero-Sutherland eigenfunctions with differing eigenvalues
\be \label{seedblockhilbert}
\psi_{e}^{(p)}(u,\bu) =  \frac{1}{(\sinh(u+\bu)\sinh(u-\bu))^{2p}} \sum_{(m,n)\in Oct^{(p)}_{\,e}} c^e_{m,n} \psi^{a_e,b_e,c_e}_{\rho_1+m, \rho_2+n}(u,\bu).
\ee
Our notation makes it explicit that $\psi^{a_e,b_e,c_e}_{\rho_1,\rho_2}$ is related to the function $F_{\rho_1,\rho_2}^{a_e, b_e, c_2}$ with $c\neq0$ (which is in turn related to the scalar conformal block). We have also introduced $\rho_1$ and $\rho_2$ which can be read off from \eqref{seedblocksolutions}. 
\paragraph{}
Alternatively, we can also directly use the gauge transformation \eqref{GaugeTrans1} to recast the coupled partial differential equations encoded in \eqref{SeedBlock-Casimir} into the form:
\bea\label{CoupledCS}
&&(\hat{\chi} \hat{\bC}^{(p)}\hat{\chi}^{-1})\cdot \hat{\chi} \vec{G}(z,\bz)=  \hat{\cH}^{(p)}\cdot \vec{\Psi}(u,\bu) = \frac{1}{2} E^l_p \vec{\Psi}(u,\bu),\nn\\
&& \hat{\chi} = {\rm diag}\left(\chi_{a_0, b_0, c_0}^{(\frac{p+2}{2})}(z,\bz),  \chi_{a_1, b_1, c_1}^{(\frac{p+2}{2})}(z,\bz), \dots, \chi_{a_e, b_e, c_e}^{(\frac{p+2}{2})}(z,\bz)\right),\nn\\
&&\hat{\cH}_{ee}^{(p)} = -\left(\frac{1}{4}\cI_2^{(\frac{p+2}{2}, e)} + \frac{1}{2}(1-c_e)^2+\left(\frac{p+2}{2}\right)\left(\frac{p+2}{2}+1-c_e\right)\right)+\frac{1}{2} \eps^p_e,\nn\\
&&\hat{\cH}_{e-1, e}^{(p)} = \frac{A_e^{(p)}}{\sinh u \sinh \bu}\left(\hat{\cL}(u,\bu)+\frac{(a-b+e-p)}{2}\right), \nn\\
&& \hat{\cH}_{e, e+1} = \frac{B_e}{\sinh u \sinh \bu} \left(\hat{\cL}(u,\bu)-\frac{(a-b+e-p)}{2}\right),\nn\\
&&\hat{\cL}(u,\bu) = \frac{\sinh^2 u\sinh^2 \bu}{2 (\sinh^2 \bu - \sinh^2 u)}\left(\coth^3 u \frac{\partial}{\partial u}-\coth^3 \bu \frac{\partial}{\partial \bu} \right)  \nn\\
&&\quad\quad\quad\quad\quad\quad\quad\quad\quad\quad\quad     -\frac{1+\sinh u\sinh\bu\cosh(u+\bu)\cosh(u-\bu)}{\sinh^2(u+\bu)\sinh^2(u-\bu)}-\frac{1}{4}.
\eea
where $\cI^{(\frac{p+2}{2},e)}_2$ indicates the specific dependence of the Hamiltonians on $(\frac{p+2}{2}, a_e, b_e, c_e)$.
One can in principle make similar linear expansion ansatz for $\Psi_e = \chi_{a_e, b_e, c_e}^{(\frac{p+2}{2})} G_e^{(p)}$ as in \eqref{seedblockhilbert}, 
in terms of Calogero-Sutherland eigenfunctions with appropriate asymptotic behaviors, and recover the same set of expansion coefficients $\{c_{mn}^{e}\}$.
As things stand, the Hilbert space seed block $\psi_{e}^{(p)}(u,\bu)$ itself \eqref{seedblockhilbert} is not an eigenfunction of  Calogero-Sutherland Hamiltonian, it remains an open question whether it can be recast into an eigenfunction of some other known quantum integrable system, it will provide a much more efficient way than iteratively deriving the various expansion coefficients here. This avenue of research could give deeper insight as to the integrable structures hidden in conformal field theories. 
\paragraph{}
We make several further comments on this result. First, the sum over the octagon of points $Oct^{(p)}_{\,e}$ seems to be an essential part of the result in \cite{Echeverri:2016dun}. Similar sums and recurrence relations can be found in \cite{Costa:2016xah} where the authors focused on an alternative radial basis for tensor structures, where the non-vanishing expansion coefficients distribute into other polygon shapes such as hexagons and their intersections. There the functions were chosen to benefit from the radial coordinates and the Gegenbauer polynomial recurrence relations. This seems in some ways the natural method for arbitrary dimensions, and we will keep it in mind for the following discussions.
\paragraph{}
This result was found for the conformal blocks with external space-time spins in four dimensions. We could ask whether a similar result holds in other dimensions. The key idea in finding the solution \eqref{seedblocksolutions} was to make an ansatz of the appropriate form; specifically, the seed blocks were assumed to be of the form of an overall function times some combination of 4d scalar conformal blocks.  In 2d, the situation is more straightforward.
In \cite{Osborn:2012vt} the conformal blocks for arbitrary spin are explicitly given by:
\bea
&&\langle \phi_1(z_1,\bz_1)\phi_2(z_2,\bz_2)\phi_3(z_3,\bz_3)\phi_4(z_4,\bz_4) \nn\\
&&= \frac{1}{z_{12}^{h_1+h_2} z_{34}^{h_3+h_4} } \frac{1}{\bz_{12}^{\bar{h}_1+\bar{h}_2} \bz_{34}^{\bar{h}_3+\bar{h}_4} }\left(\frac{z_{24}}{z_{14}}\right)^{h_{12}}\left(\frac{z_{14}}{z_{13}}\right)^{h_{34}}\left(\frac{\bz_{24}}{\bz_{14}}\right)^{\bar{h}_{12}}\left(\frac{\bz_{14}}{\bz_{13}}\right)^{\bar{h}_{34}} \mathcal{F}_{1234}(z,\bz),\nn\\
\eea
where $h_i +\bar{h}_i=\Del_i$, $\bar{h}_i-\bar{h}_i = s_i$ and $s_i$ is the 2d spin.
One should note that $\{z_i\}$ here denote the complex 2d space-time coordinates, while the summation over all the exchange operators $\cO_{h,\bar{h}}$ is given by
\be\label{F1234}
\mathcal{F}_{1234}(z,\bz) = \sum_{h,\bar{h}} a_{h,\bar{h}}F_{h,\bar{h}}(z,\bz), \quad z = \frac{z_{12}z_{34}}{z_{13}z_{24}},~~ \bz = \frac{\bz_{12}\bz_{34}}{\bz_{13}\bz_{24}},
\ee
and
\be\label{Def:Fzbz}
F_{h,\bar{h}}(z,\bz)=z^h \bz^{\bar{h}} 
{}_2 F_1({h}+{h}_{21},{h}+{h}_{34},2{h},z)
{}_2 F_1(\bar{h}+\bar{h}_{21},\bar{h}+\bar{h}_{34},2\bar{h},\bz).
\ee
We can regard the summation \eqref{F1234} as the 2d analogue of decomposition into independent tensor structures. 
For given external spin configurations, the expansion coefficients $a_{h,\bar{h}}$ can be calculated from the contribution of a conformal primary operator $\mathcal{O}$ to the $\phi_1\phi_2$ OPE:
\be
\phi_1(z_1,\bz_1)\phi_2(z_2,\bz_2) \sim \frac{C_{12\mathcal{O}} z_{12}^h  \bz_{12}^{\bar{h}}}
{z_{12}^{h_1+h_2}  \bz_{12}^{\bar{h}_1+\bar{h}_2}}
 {}_1F_1({h}+{h}_{12},2{h}, z_{12}\partial_{z_2})
{}_1F_1(\bar{h}+\bar{h}_{12}, 2\bar{h}, \bz_{12}\partial_{\bz_2})\mathcal{O}(z_2,\bz_2).
\ee
For real scalars $\phi_i$, $\bar{h}_i=h_i$ and the functions $F_{h,\bar{h}}$ are scalar conformal blocks. 
While in general the $z$ and $\bz$ dependent pieces in \eqref{Def:Fzbz} are eigenfunctions of $D_z(h_{21}, h_{34}, h)$ and $D_\bz(\bar{h}_{21}, \bar{h}_{34},\bar{h})$ respectively,
with the eigenvalues $h$ and $\bar{h}$ (cf. \eqref{eigenDz}).
By acting with the gauge transformation \eqref{GaugeTrans1}, these conformal blocks transform to eigenfunctions of the P{\"o}schl-Teller Hamiltonian:
\be
\Psi(z(u),\bz(\bu))=\psi(z(u))\bar{\psi}(\bz(\bu)),\quad \psi(z(u)) \sim z(u)^{h-\frac12} (z(u)-1)^{\frac{h_{21}+h_{34}}{2}+\frac14}{}_2 F_1(h+h_{21},h+h_{34}, 2h , z(u)).
\ee
similarly for $\bar{\psi}(\bz(\bu))$ with $(h, h_{21}, h_{34}) \to (\bar{h}, \bar{h}_{21}, \bar{h}_{34})$ as discussed in \cite{Isachenkov:2016gim}. 
Thus in 2d we once again find that the Hilbert space seed is given by some linear combinations of gauge-transformed scalar conformal blocks.
\paragraph{}
The question of three or other odd dimensions is more difficult, as scalar conformal blocks in these dimensions do not have a simple closed form. In \cite{Hogervorst:2016hal}, a formula is obtained for 3d conformal blocks in terms of infinitely many 2d conformal blocks. We conjecture that a similar understanding holds, and the Hilbert space seed blocks in 3d are related in a similar manner to 3d scalar conformal blocks. 
\paragraph{}
In other arbitrary dimensions, we can only speculate. In \cite{Costa:2016xah} the space-time dimension enters as a parameter. It may be possible that spinning conformal blocks in fractional dimensions are always expressible in terms of a basis of scalar conformal blocks in the same fractional dimension. Another direction to consider is whether a similar result holds in 6d. We have analytic expressions for 6d scalar conformal blocks in terms of sum of hypergeometric functions. Beginning with a similar ansatz and using hypergeometric recurrence relations, it is possible that the Hilbert space seed blocks should be an overall function times a linear combination of Calogero-Sutherland eigenfunctions with differing eigenvalues.

\subsection*{Acknowledgement}
This work was supported in part by Ministry of Science and Technology through the grant 104-2112-M-002 -004 -MY3, Center for Theoretical Sciences at National Taiwan University, and National Center for Theoretical Sciences. We are very grateful to Mikail Isachenkov and Volker Schomerus for valuable comments.

\appendix
\section{Embedding and Super-embedding formalisms}\label{appendix}\label{App:Embedding}
\paragraph{}
A convenient formalism for investigating conformal blocks and the quadratic conformal (and ultimately superconformal) Casimir is the embedding space formalism. 
Here we present only the absolutely relevant details, and direct the readers to the references \cite{Embedding-refs} for more detailed information.
\paragraph{}
The generators of the $d$-dimensional conformal group can be identified with the generators of the Lorentz group $SO(d+1,1)$ of $\bbR^{d+1,1}$,
such that the $d$-dimensional conformal transformations acting on $\bbR^d$ can be realized linearly as the Lorentz transformation acting on the embedding space $\mathbb{R}^{d+1,1}$. 
Let us define coordinates $P^m=(X^+,X^-,X^\mu)$ with inner product:
\be
P \cdot P = \eta_{AB}P^A P^B = -X^+X^- + X_\mu X^\mu,\quad   \quad A, B =0, \dots, d+1.
\ee
In order to recover $d$-dimensional conformal from $d+2$ dimensional Lorentz transformation, we first restrict to the null-cone $P^2=0$, and Euclidean space is recovered by making the identification $P\sim\lambda P, \lambda>0$. By imposing the gauge condition $X^+=1$, we have 
\be
P_{AB}\equiv-2P_1\cdot P_2 = (x_1-y_2)^2,
\ee
Scalar primary operators $\Phi(P)$ can be understood as homogenous fields on the null-cone
\be
\Phi(\lambda P) = \lambda^{-\Delta}\Phi(P),
\ee
where 
$\Phi(P)\equiv (X^+)^{-\Delta}\phi(X^\mu/X^+)$
for a primary scalar $\phi(x)$ with dimension $\Delta$. Any correlator involving $\Phi(P)$ must respect this homogeneity, so that the decomposition (\ref{4pt-function-Para}) can be written as 
\be \label{4pt-function-embed}
\langle \Phi_1(P_1)\Phi_2(P_2)\Phi_3(P_3)\Phi_4(P_4) \rangle = \frac{1}{(P_{12}^2)^{\frac{1}{2} (\Del_1+\Del_2)} (P_{34}^2)^{\frac{1}{2} (\Del_3+\Del_4)}} \left(\frac{P_{14}^2}{P_{24}^2}\right)^{a} \left(\frac{P_{14}^2}{P_{13}^2}\right)^{b} F(\ru, \rv).
\ee
The advantage of writing correlation functions in embedding space coordinates is that the action of conformal Casimir operators is straightforward. 
For the quadratic conformal Casimir, this action is an eigenvalue equation:
\be
\frac12 (J_{1,AB}+J_{2,AB})(J_{1}^{AB}+J_{2}^{AB}) \langle \Phi_1(P_1)\Phi_2(P_2)\Phi_3(P_3)\Phi_4(P_4) \rangle = 
\frac{1}{2}c_2(\Delta,\ell) \langle \Phi_1\Phi_2\Phi_3\Phi_4 \rangle,
\ee
where $J_i$ is the Lorentz generator acting on $P_i$ given by
\be\label{lorentz-gen}
J_{i, AB} \equiv i\left( P_{i, A} \frac{\partial}{\partial P^B_i}  - P_{i,B} \frac{\partial}{\partial P^A_i} \right).
\ee
On the one hand, we can find the conformal Casimir eigenvalue by considering the quadratic Casimir operator in terms of the original conformal group generators $D, P, K, M$. On the other hand, we can apply the operator \eqref{lorentz-gen} to our ansatz \eqref{4pt-function-embed} to find an explicit partial differential equation \cite{Dolan:2003hv, Dolan:2011dv}.
\paragraph{}
We generalize to superembedding space formalism \cite{Lemos:2015awa,N=1 Blocks, Goldberger Papers} in order to obtain the superblock as an eigenfunction of the superconformal Casimir operator in $d=4$ dimensions. As pointed out in \cite{N=1 Blocks}, this superspace construction can describe multiplets whose superconformal primary is invariant under a nonabelian $R$-symmetry group. This will clearly include all $\mathcal{N}=1$ multiplets, as well as some multiplets in extended SUSY. 
The analogue of the bosonic $P$ coordinate is the supertwistor:
\be
Z_A =  \left( \begin{matrix}
Z_\alpha \\
Z^{\dot\alpha} \\
Z_i \\
\end{matrix} \right) \in \mathbb{C}^{4|\mathcal{N}},
\ee
where $Z_\alpha,Z^{\dot\alpha}$ are bosonic components and $Z_i$ are $\mathcal{N}$ fermionic components, with $\alpha,\dot{\alpha}=1,\dots,4$ and $i=1,\dots,\mathcal{N}$.
We also have dual supertwistors $\bar{Z}\equiv Z^\dagger\Omega$ with components
\be
\bar{Z}^A = \left(\begin{matrix}   \bar{Z}^\alpha & \bar{Z}_{\dot\alpha}  &  \bar{Z}^i.
\end{matrix}\right)
\ee
The matrix $\Omega$ defines the inner product $\langle Z_1,Z_2  \rangle = Z^\dagger_1\Omega Z_2$ and is given by
\be \Omega \equiv \left(\begin{matrix}
0 & \delta^{\dot\beta}_{\dot\alpha} & 0 \\
\delta_{\beta}^{\alpha} & 0 & 0 \\
0 & 0 & \delta^{i}_{j} \\
\end{matrix}\right).
\ee
As discussed in \cite{N=1 Blocks}, these super- and dual supertwistors have $GL(2,\mathbb{C})$ gauge redundancies that act as a change of basis:
\be
Z_A^a \sim Z_A^b g_b^a, \;\;\;\;\;\; g_b^a \in GL(2,\mathbb{C}),
\ee
and similarly for $\bar{Z}$.
Because physical quantities are independent of these redundancies, we introduce bitwistors
\be
X_{AB}\equiv Z^a_A Z^b_B \epsilon_{ab}, \;\;\;\;\;\; \bar{X}^{AB}\equiv \bar{Z}^{\dot{a}A} \bar{Z}^{\dot{b}B} \epsilon_{\dot{a}\dot{b}}
\ee
which are well-defined up to rescaling
\be
(X,\bar{X}) \sim (\lambda X,\bar{\lambda}\bar{X}), \;\;\;\;\;\;\lambda=\det g, \bar{\lambda}=\det \bar{g}.
\ee
The superembedding coordinates are defined to satisfy: 
\begin{gather}
(X,\bar{X}) \sim (\lambda X,\bar{\lambda}\bar{X}), ~~ X_{AB}=-(-1)^{p_A p_B}X_{BA},
\end{gather}
where $p_A=0$ when $A=\alpha$ and $p_A=1$ when $A=i$. 
We also use the superconformal invariants
\begin{gather}
\langle 1\bar{2}  \rangle = \bar{X}_2^{AB}X_{1,BA}, ~~ \langle 1\bar{2} 3 \bar{4} \rangle = \bar{X}_4^{AB}X_{3,BC} \bar{X}_2^{CD}X_{1,DA}(-1)^{p_C}.
\end{gather}
We represent chiral fields by the holomorphic function $\Phi(X)$ and anti-chiral fields by the anti-holomorphic $\Phi(\bar{X})$. We consider four point correlation function of the form:
\be \label{fourpointsuper}
\langle {\Phi}({X}_1) \bar{\Phi}(\bar{X}_2) {\Phi}({X}_3) \bar{\Phi}(\bar{X}_4)
 \rangle = 
  \frac{1}{\langle 1\bar{2}  \rangle^\Delta_\Phi \langle 3\bar{4} \rangle^\Delta_\Phi}G(\ru,\rv).
\ee
The variables $\ru$ and $\rv$ are the superconformal analogues of the usual bosonic cross ratios 
\be
\frac{ \langle 1\bar{2}3\bar{4} \rangle }{\langle 1 \bar{4} \rangle \langle 3 \bar{2} 3 \rangle} = \frac{\ru+\rv-1}{4\rv}, \quad \frac{\langle 1\bar{2}  \rangle \langle 3\bar{4} \rangle}{\langle 1 \bar{4} \rangle \langle 3 \bar{2} 3 \rangle}=\frac{\ru}{\rv}.
\ee
The superembedding space analogue to the Lorentz generators $J_{AB}$ are given by 
\be
L_A^B \equiv Z^a_A \frac{\partial}{\partial Z^a_B}-\bar{Z}^{\dot{a}B} \frac{\partial}{\partial \bar{Z}^{\dot{a}A}}(-1)^{p_Ap_B}.
\ee
It can be difficult to keep track of the sign factors coming from the grading of the components of $Z,\bar{Z}$.
We follow the trick of \cite{N=1 Blocks} and pretend that these quantities are purely bosonic, transforming under $SL(n)$. This $SL(n)$ invariance guarantees that any $n$-dependence comes from traces of the identity matrix. That is, we perform computations pretending $Z,\bar{Z}\in\mathcal{C}^{n\times2}$ and set $n=4-\mathcal{N}$ to recover the result for the superconformal group $SU(2,2|\mathcal{N})$.
\paragraph{}
Thus in order to calculate the superconformal quadratic Casimir, we compute the Casimir operator for $SL(n)$:
\be 
C_n = L_A^B - \frac1n L_A^A L_B^B.
\ee
Acting on the four point correlation function \eqref{fourpointsuper}, one finds a differential equation for the superconformal block for various values of $\mathcal{N}$. This equation
relates to the Casimir equation for scalar fields with different scaling dimensions \cite{Dolan:2003hv}, and relating these differential equations we find that we can express certain superconformal blocks in terms of the standard bosonic conformal blocks. Although these superembedding space calculations were performed for $d=4$ dimensions, we can conjecture a natural generalization to other dimensions.
\bibliographystyle{sort}

\end{document}